\documentclass[a4paper,10pt]{article}

\usepackage{epsf}
\usepackage{graphicx}    % Include figure files

\newcommand{\be}[1]{\begin{equation}\label{#1}}
\newcommand{\ee}{\end{equation}}
\newcommand{\bea}{\begin{eqnarray}}
\newcommand{\eea}{\end{eqnarray}}

\def\gsim{ \lower .75ex \hbox{$\sim$} \llap{\raise .27ex \hbox{$>$}} }
\def\lsim{ \lower .75ex \hbox{$\sim$} \llap{\raise .27ex \hbox{$<$}} }

\renewcommand{\markright}{\markright{\thepage}}

\begin{document}

\begin{titlepage}

%\begin{flushright}
%astro-ph/0609597
%\end{flushright}

\vspace{5mm}

\begin{center}

{\Large \bf Dynamical vacuum energy, holographic quintom, and the
reconstruction of scalar-field dark energy}

\vspace{10mm}

{\large Xin Zhang}

\vspace{5mm} {\em %CCAST (World Laboratory), P.O.Box 8730, Beijing
%100080, People's Republic of China\\
Institute of Theoretical Physics, Chinese Academy of Sciences,
P.O.Box 2735, Beijing 100080, People's Republic
of China\\
Interdisciplinary Center of Theoretical Studies, Chinese Academy of
Sciences, P.O.Box 2735, Beijing 100080, People's Republic of China}

\end{center}

\vspace{5mm}
\begin{abstract}
When taking the holographic principle into account, the vacuum
energy will acquire dynamical property that its equation of state is
evolving. The current available observational data imply that the
holographic vacuum energy behaves as quintom-type dark energy. We
adopt the viewpoint of that the scalar field models of dark energy
are effective theories of an underlying theory of dark energy. If we
regard the scalar field model as an effective description of such a
holographic vacuum theory, we should be capable of using the scalar
field model to mimic the evolving behavior of the dynamical vacuum
energy and reconstructing this scalar field model according to the
fits of the observational dataset. We find the generalized ghost
condensate model is a good choice for depicting the holographic
vacuum energy since it can easily realize the quintom behavior. We
thus reconstruct the function $h(\phi)$ of the generalized ghost
condensate model using the best-fit results of the observational
data.
%\noindent PACS numbers: 95.36.+x, 98.80.Es, 98.80.-k
\end{abstract}

\end{titlepage}

\newpage

\setcounter{page}{2}

Many cosmological experiments, such as observations of large scale
structure \cite{LSS}, searches for type Ia supernovae \cite{SN}, and
measurements of the cosmic microwave background anisotropy
\cite{CMB}, all indicate that the expansion of the universe is
undergoing cosmic acceleration at the present time. This cosmic
acceleration is viewed as due to a mysterious dominant component,
dark energy, with negative pressure. The combined analysis of
cosmological observations suggests that the universe is spatially
flat, and consists of about $70\%$ dark energy, $30\%$ dust matter
(cold dark matter plus baryons), and negligible radiation. Although
we can affirm that the ultimate fate of the universe is determined
by the feature of dark energy, the nature of dark energy as well as
its cosmological origin remain enigmatic at present. The most
obvious theoretical candidate of dark energy is the cosmological
constant $\Lambda$ (vacuum energy) \cite{Einstein:1917,cc} which has
the equation of state $w=-1$. However, as is well known, there are
two difficulties arise from the cosmological constant scenario,
namely the two famous cosmological constant problems --- the
``fine-tuning'' problem and the ``cosmic coincidence'' problem
\cite{coincidence}. The fine-tuning problem asks why the vacuum
energy density today is so small compared to typical particle
scales. The vacuum energy density is of order $10^{-47} {\rm
GeV}^4$, which appears to require the introduction of a new mass
scale 14 or so orders of magnitude smaller than the electroweak
scale. The second difficulty, the cosmic coincidence problem, says
``Since the energy densities of vacuum energy and dark matter scale
so differently during the expansion history of the universe, why are
they nearly equal today''? To get this coincidence, it appears that
their ratio must be set to a specific, infinitesimal value in the
very early universe.

An alternative proposal for dark energy is the dynamical dark energy
scenario. The cosmological constant puzzles may be better
interpreted by assuming that the vacuum energy is canceled to
exactly zero by some unknown mechanism and introducing a dark energy
component with a dynamically variable equation of state. The
dynamical dark energy proposal is often realized by some scalar
field mechanism which suggest that the energy form with negative
pressure is provided by a scalar field evolving down a proper
potential. Actually, this mechanism is enlightened to a great extent
by the inflationary cosmology. As we have known, the occurrence of
the current accelerating expansion of the universe is not the first
time for the expansion history of the universe. There is significant
observational evidence strongly supports that the universe underwent
an early inflationary epoch, over sufficiently small time scales,
during which its expansion rapidly accelerated under the driven of
an ``inflaton'' field which had properties similar to those of a
cosmological constant. The inflaton field, to some extent, can be
viewed as a kind of dynamically evolving dark energy. Hence, the
scalar field models involving a minimally coupled scalar field are
proposed, inspired by inflationary cosmology, to construct
dynamically evolving models of dark energy. The only difference
between the dynamical scalar-field dark energy and the inflaton is
the energy scale they possess. So far, a host of scalar-field dark
energy models have been studied, including quintessence
\cite{quintessence}, K-essence \cite{kessence}, tachyon
\cite{tachyon}, phantom \cite{phantom}, ghost condensate
\cite{ghost1,ghost2} and quintom \cite{quintom} etc.. Generically,
there are two points of view on the scalar-field models of dynamical
dark energy. One viewpoint regards the scalar field as a fundamental
field of the nature. The nature of dark energy is, according to this
viewpoint, completely attributed to some fundamental scalar field
which is omnipresent in supersymmetric field theories and in
string/M theory. The other viewpoint supports that the scalar field
model is an effective description of an underlying theory of dark
energy. On the whole, it seems that the latter is the mainstream
point of view. Since we regard the scalar field model as an
effective description of an underlying theory of dark energy, a
question arises asking: What is the underlying theory of the dark
energy? Of course, hitherto, this question is far beyond our present
knowledge, because that we can not entirely understand the nature of
dark energy before a complete theory of quantum gravity is
established. However, although we are lacking a quantum gravity
theory today, we still can make some attempts to probe the nature of
dark energy according to some principles of quantum gravity. The
holographic dark energy model is just an appropriate example, which
is constructed in the light of the holographic principle of quantum
gravity theory. That is to say, the holographic dark energy model
possesses some significant features of an underlying theory of dark
energy.

The distinctive feature of the cosmological constant or vacuum
energy is that its equation of state is always exactly equal to
$-1$. However, when considering the requirement of the holographic
principle originating from the quantum gravity speculation, the
vacuum energy will become dynamically evolving dark energy.
Actually, the dark energy problem may be in principle a problem
belongs to quantum gravity \cite{Witten:2000zk}. In the classical
gravity theory, one can always introduce a cosmological constant to
make the dark energy density be an arbitrary value. However, a
complete theory of quantum gravity should be capable of making the
property of dark energy, such as the equation of state, be
determined definitely and uniquely \cite{Witten:2000zk}. Currently,
an interesting attempt for probing the nature of dark energy within
the framework of quantum gravity is the so-called ``holographic dark
energy'' proposal
\cite{Cohen:1998zx,Horava:2000tb,Hsu:2004ri,Li:2004rb}. It is well
known that the holographic principle is an important result of the
recent researches for exploring the quantum gravity (or string
theory) \cite{holoprin}. This principle is enlightened by
investigations of the quantum property of black holes. Simply
speaking, in a quantum gravity system, the conventional local
quantum field theory will break down. The reason is rather simple:
For a quantum gravity system, the conventional local quantum field
theory contains too many degrees of freedom, and such many degrees
of freedom will lead to the formation of black hole so as to break
the effectiveness of the quantum field theory.

For an effective field theory in a box of size $L$, with UV cut-off
$\Lambda_c$ the entropy $S$ scales extensively, $S\sim
L^3\Lambda_c^3$. However, the peculiar thermodynamics of black hole
\cite{bh} has led Bekenstein to postulate that the maximum entropy
in a box of volume $L^3$ behaves nonextensively, growing only as the
area of the box, i.e. there is a so-called Bekenstein entropy bound,
$S\leq S_{BH}\equiv\pi M_P^2L^2$. This nonextensive scaling suggests
that quantum field theory breaks down in large volume. To reconcile
this breakdown with the success of local quantum field theory in
describing observed particle phenomenology, Cohen et al.
\cite{Cohen:1998zx} proposed a more restrictive bound -- the energy
bound. They pointed out that in quantum field theory a short
distance (UV) cut-off is related to a long distance (IR) cut-off due
to the limit set by forming a black hole. In other words, if the
quantum zero-point energy density $\rho_{\rm de}$ is relevant to a
UV cut-off, the total energy of the whole system with size $L$
should not exceed the mass of a black hole of the same size, thus we
have $L^3\rho_{\rm de}\leq LM_P^2$. This means that the maximum
entropy is in order of $S_{BH}^{3/4}$. When we take the whole
universe into account, the vacuum energy related to this holographic
principle \cite{holoprin} is viewed as dark energy, usually dubbed
holographic dark energy. The largest IR cut-off $L$ is chosen by
saturating the inequality so that we get the holographic dark energy
density
\begin{equation}
\rho_{\rm de}=3c^2M_P^2L^{-2}~,\label{de}
\end{equation} where $c$ is a numerical constant, and $M_P\equiv 1/\sqrt{8\pi
G}$ is the reduced Planck mass. Hereafter, we will use the unit
$M_P=1$ for convenience. If we take $L$ as the size of the current
universe, for instance the Hubble scale $H^{-1}$, then the dark
energy density will be close to the observed data. However, Hsu
\cite{Hsu:2004ri} pointed out that this yields a wrong equation of
state for dark energy. Li \cite{Li:2004rb} subsequently proposed
that the IR cut-off $L$ should be taken as the size of the future
event horizon
\begin{equation}
R_{\rm eh}(a)=a\int_t^\infty{dt'\over a(t')}=a\int_a^\infty{da'\over
Ha'^2}~.\label{eh}
\end{equation} Then the problem can be solved nicely and the
holographic dark energy model can thus be constructed successfully.
The holographic dark energy scenario may provide simultaneously
natural solutions to both dark energy problems as demonstrated in
Ref.\cite{Li:2004rb}. The holographic dark energy model has been
tested and constrained by various astronomical observations
\cite{obs1,obs2,obs3}. For other extensive studies, see e.g.
\cite{holoext}.

Consider now a spatially flat FRW (Friedmann-Robertson-Walker)
universe with matter component $\rho_{\rm m}$ (including both baryon
matter and cold dark matter) and holographic dark energy component
$\rho_{\rm de}$, the Friedmann equation reads
\begin{equation}
3H^2=\rho_{\rm m}+\rho_{\rm de}~,
\end{equation} or equivalently,
\begin{equation}
{H^2\over H_0^2}=\Omega_{\rm m}^0a^{-3}+\Omega_{\rm de}{H^2\over
H_0^2}~.\label{Feq}
\end{equation}
Note that we always assume spatial flatness throughout this paper as
motivated by inflation. Combining the definition of the holographic
dark energy (\ref{de}) and the definition of the future event
horizon (\ref{eh}), we derive
\begin{equation}
\int_a^\infty{d\ln a'\over Ha'}={c\over Ha\sqrt{\Omega_{\rm
de}}}~.\label{rh}
\end{equation} We notice that the Friedmann
equation (\ref{Feq}) implies
\begin{equation}
{1\over Ha}=\sqrt{a(1-\Omega_{\rm de})}{1\over
H_0\sqrt{\Omega_m^0}}~.\label{fri}
\end{equation} Substituting (\ref{fri}) into (\ref{rh}), one
obtains the following equation
\begin{equation}
\int_x^\infty e^{x'/2}\sqrt{1-\Omega_{\rm de}}dx'=c
e^{x/2}\sqrt{{1\over\Omega_{\rm de}}-1}~,
\end{equation} where $x=\ln a$. Then taking derivative with respect to $x$ in both
sides of the above relation, we get easily the dynamics satisfied by
the dark energy, i.e. the differential equation about the fractional
density of dark energy,
\begin{equation}
\Omega'_{\rm de}=-(1+z)^{-1}\Omega_{\rm de}(1-\Omega_{\rm
de})\left(1+{2\over c}\sqrt{\Omega_{\rm de}}\right),\label{deq}
\end{equation}
where the prime denotes the derivative with respect to the redshift
$z$. This equation describes behavior of the holographic dark energy
completely, and it can be solved exactly \cite{Li:2004rb}. From the
energy conservation equation of the dark energy, the equation of
state of the dark energy can be given \cite{Li:2004rb}
\begin{equation}
w=-1-{1\over 3}{d\ln\rho_{\rm de}\over d\ln a}=-{1\over 3}(1+{2\over
c}\sqrt{\Omega_{\rm de}})~.\label{w}
\end{equation} Note that the formula
$\rho_{\rm de}={\Omega_{\rm de}\over 1-\Omega_{\rm de}}\rho_{\rm
m}^0a^{-3}$ and the differential equation of $\Omega_{\rm de}$
(\ref{deq}) are used in the second equal sign. It can be seen
clearly that the equation of state of the holographic dark energy
evolves dynamically and satisfies $-(1+2/c)/3\leq w\leq -1/3$ due to
$0\leq\Omega_{\rm de}\leq 1$. Hence, we see clearly that when taking
the holographic principle into account the vacuum energy becomes
dynamically evolving dark energy. The parameter $c$ plays a
significant role in this model. If one takes $c=1$, the behavior of
the holographic dark energy will be more and more like a
cosmological constant with the expansion of the universe, such that
ultimately the universe will enter the de Sitter phase in the far
future. As is shown in \cite{Li:2004rb}, if one puts the parameter
$\Omega_{\rm de}^0=0.73$ into (\ref{w}), then a definite prediction
of this model, $w_0=-0.903$, will be given. On the other hand, if
$c<1$, the holographic dark energy will exhibit appealing behavior
that the equation of state crosses the ``cosmological-constant
boundary'' (or ``phantom divide'') $w=-1$ during the evolution. This
kind of dark energy is referred to as ``quintom'' \cite{quintom}
which is slightly favored by current observations
\cite{cross1,cross2}. If $c>1$, the equation of state of dark energy
will be always larger than $-1$ such that the universe avoids
entering the de Sitter phase and the Big Rip phase. Hence, we see
explicitly, the value of $c$ is very important for the holographic
dark energy model, which determines the feature of the holographic
dark energy as well as the ultimate fate of the universe.

The holographic dark energy model has been tested and constrained by
various astronomical observations \cite{obs1,obs2,obs3}. In recent
works \cite{obs1,obs2}, it has been explicitly shown that regarding
the observational data including type Ia supernovae (SN), cosmic
microwave background (CMB), baryon acoustic oscillation (BAO), and
the X-ray gas mass fraction of galaxy clusters (X-ray gas), the
holographic dark energy behaves like a quintom-type dark energy.
This indicates that the numerical parameter $c$ in the model is less
than 1. The main constraint results are summarized as follows:
\begin{enumerate}

\item Using only the SN data to constrain the holographic dark
energy model, we get the fit results: $c=0.21^{+0.41}_{-0.12}$,
$\Omega_{\rm m}^0=0.47^{+0.06}_{-0.15}$, with the minimal chi-square
corresponding to the best fit $\chi^2_{\rm min}=173.44$ \cite{obs1}.
In this fitting, the 157 gold data points listed in Riess et al.
\cite{Riess:2004nr}, including 14 high redshift SN (gold) data from
the HST/GOODS program, have been used to constrain the model. For
the plot of the confidence level contours of $68\%$, $95\%$ and
$99\%$ in the $(c, \Omega_{\rm m}^0)$ plane see Fig.2 of
Ref.\cite{obs1}. We notice in this figure that the current SN Ia
data do not strongly constrain the parameters $\Omega_{\rm m}^0$ and
$c$ (in $2\sigma$), in particular $c$, in the considered ranges.
According to the best fit result, the value of $c$ is significant
smaller than 1, resulting in that the present equation-of-state of
dark energy is $w_0=-2.64$, which seems not a reasonable result. The
present deceleration parameter is $q_0=-1.60$ and the
deceleration/acceleration happens at $z_T=0.27$. Of course, other
observations may impose further constraints. For instance, the CMB
and LSS data can provide us with useful complements to the SN data
for constraining cosmological models. It has been shown in
Ref.\cite{obs1} that it is very important to find other
observational quantities irrelevant to $H_0$ as complement to SN Ia
data. Such suitable data can be found in the probes of CMB and BAO.

\item Combining the information from SN Ia \cite{Riess:2004nr},
CMB \cite{CMB} and BAO \cite{BAO}, the fitting for the holographic
dark energy model gives the parameter constraints in 1 $\sigma$:
$c=0.81^{+0.23}_{-0.16}$, $\Omega_{\rm m}^0=0.28\pm 0.03$, with
$\chi_{\rm min}^2=176.67$ \cite{obs1}. For the confidence contour
plot see Fig.6 of Ref.\cite{obs1}. We see clearly that a great
progress has been made when we perform a joint analysis of SN Ia,
CMB, and BAO data. Note that the best fit value of c is also less
than 1, though in $1\sigma$ range it can slightly larger than 1. For
the SN+CMB+BAO joint analysis, according to the best fit, we derive
that the deceleration parameter $q$ has a value of $q_0=-0.61$ at
present. The transition from deceleration to acceleration
($q(z_T)=0$) occurs at a redshift of $z_T=0.63$. The
equation-of-state parameter $w$ is slightly smaller than $-1$ at
present, $w_0=-1.03$. These results look very rational, and also
favor a quintom-type holographic dark energy.

\item Although the SN+CMB+BAO joint analysis provides a fairly good
constraint result for the holographic dark energy model, it is,
however, necessary to test dark energy model and constrain the
parameters using as many techniques as possible. Different tests
might provide different constraints on the parameters of the model,
and a comparison of results determined from different methods allows
us to make consistency checks. Therefore, the X-ray gas mass
fraction of rich clusters, as a function of redshift, has also been
used to constrain the holographic dark energy model \cite{obs2}. The
main results, i.e. the 1 $\sigma$ fit values for $c$ and
$\Omega_{\rm m}^0$ are: $c=0.61^{+0.45}_{-0.21}$ and
$\Omega_m^0=0.24^{+0.06}_{-0.05}$, with the best-fit chi-square
$\chi_{\rm min}^2=25.00$ \cite{obs2}. The plot of 1, 2 and 3
$\sigma$ confidence level contours in the $(c, \Omega_{\rm m}^0)$
plane is shown in Fig.1 of Ref.\cite{obs2}. We notice that the fit
value of $c$ is less than 1 in 1 $\sigma$ range, though it can be
slightly larger than 1. This implies that according to the $f_{\rm
gas}$ constraints the holographic dark energy basically behaves as a
quintom-type dark energy in 1 $\sigma$ range. At the best-fit, we
derive that the equation of state of dark energy $w$ has a value of
$w_0=-1.29$ and the deceleration parameter $q$ has a value of
$q_0=-0.97$ at present. The typical characteristic of the
quintom-type dark energy is that the equation of state can cross
$-1$. For this case, the crossing behavior $(w(z_C)=-1)$ happens at
a redshift of $z_C=0.62$. In addition, the transition from
deceleration to acceleration $(q(z_T)=0)$ occurs at the redshift
$z_T=0.70$.

\item Finally, we shall mention the results of a Monte Carlo
simulation of the SNAP mission for analyzing the holographic dark
energy model. To find the expected precision of the SNAP, one must
assume a fiducial model, and then simulate the experiment assuming
it as a reference model. We can use SNAP specifications to construct
mock SN catalogues. The best fit values for the model parameters are
$c=0.92$ and $\Omega_{\rm m}^0=0.23$, when assuming a $\Lambda$CDM
model as fiducial model with $\Omega_{\rm m}^0=0.27$ and $h=0.71$.
For the predicted confidence level contours in the $(c, \Omega_{\rm
m}^0)$ plane for this simulation see Fig.9 of Ref.\cite{obs1}. We
notice with interest that the precision type Ia supernova
observations will still support a quintom type holographic dark
energy.

\end{enumerate}
On the whole, through the various observational constraints, we
conclude that the parameter $c$ is smaller than 1 so as to make the
holographic dark energy behave as a quintom-type dark energy. We
refer to this case as ``holographic quintom''. In the light of the
best fit results of various observational data analyses, we plot in
Fig.1 the evolutions of the equation of state of dark energy
component.

%%%%%%%%%%%%%%%%%%%%%%%%%%%%%%%%%%%%%%%%%%%%%%%%%%%%%%%%%%%%%%%%%%
\begin{figure}[htbp]
\begin{center}
\includegraphics[scale=1.2]{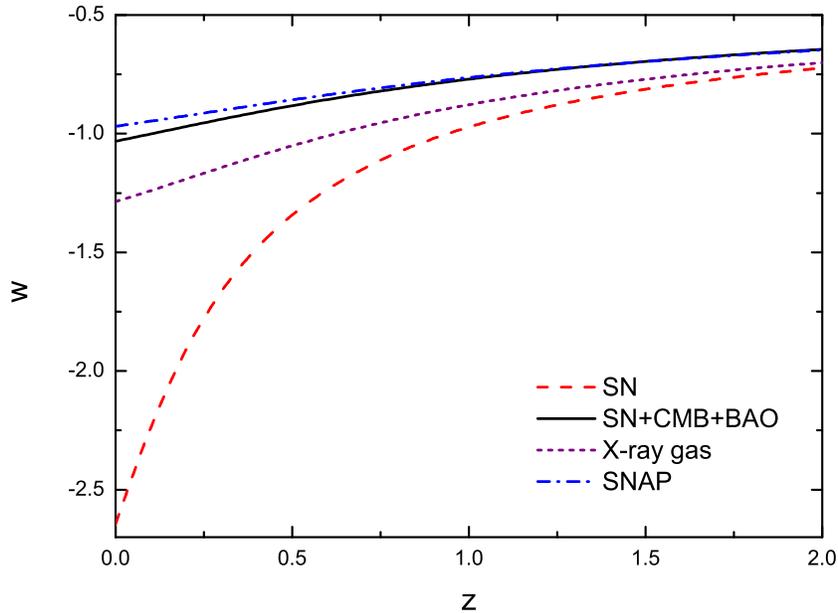}
\caption[]{\small The evolutions of the equation of state of
holographic dark energy. Here we use the best fit results of the
type Ia supernovae, the joint analysis of SN+CMB+BAO, the X-ray gas
mass fraction of galaxy clusters, and the Monte Carlo simulation of
SNAP mission, respectively. In the concrete, $c=0.21$ and
$\Omega_{\rm m}^0=0.47$ for SN; $c=0.81$ and $\Omega_{\rm m}^0=0.28$
for SN+CMB+BAO; $c=0.61$ and $\Omega_{\rm m}^0=0.24$ for X-ray gas;
$c=0.92$ and $\Omega_{\rm m}^0=0.23$ for SNAP. }
\end{center}
\end{figure}
%%%%%%%%%%%%%%%%%%%%%%%%%%%%%%%%%%%%%%%%%%%%%%%%%%%%%%%%%%%%%%%%%%%

As has been analyzed above, the holographic dark energy scenario
reveals the dynamical nature of the vacuum energy. When taking the
holographic principle into account, the vacuum energy density will
evolve dynamically. In particular, the analysis of the observational
data indicates that the holographic vacuum energy is likely to
behave as quintom dark energy. On the other hand, as has already
mentioned, the scalar field dark energy models are often viewed as
effective description of the underlying theory of dark energy.
However, the underlying theory of dark energy can not be achieved
before a complete theory of quantum gravity is established. We can,
nevertheless, speculate on the underlying theory of dark energy by
taking some principles of quantum gravity into account. The
holographic dark energy model is no doubt a tentative in this way.
We are now interested in that if we assume the holographic vacuum
energy scenario as the underlying theory of dark energy, how the
scalar field model can be used to effectively describe it.

It should be pointed out that the quintom type dark energy whose
equation-of-state crosses the cosmological-constant boundary
($w=-1$) can not be realized by an ordinary minimally coupled scalar
field [$p=X-V(\phi)$].\footnote{The crossing to the phantom region
($w<-1$) can often be realized in terms of a two-field system with a
phantom field and an ordinary scalar field (quintessence). But in
this paper, we only focus on the single-field model.} This
transition of crossing $w=-1$ can occur for the Lagrangian density
$p(\phi, X)$ in which $\partial{p}/\partial{X}$ changes sign from
positive to negative, but we require nonlinear terms in $X$ to
realize the $w=-1$ crossing \cite{nl,general}. It has been shown in
Ref.\cite{general} that a simple one-field model, generalized ghost
condensate, can easily realize the crossing cosmological-constant
boundary. We shall use this scalar field model to effectively
describe the holographic quintom vacuum energy, and perform the
reconstruction of such a scalar model. For the reconstruction of
dark energy models, see e.g.
\cite{general,Saini:1999ba,simplescalar,Boisseau:2000pr,Szydlowski:2003cf,Zhang:2006av,Guo:2005at,Zhao:2006mp}.

First, let us consider the Lagrangian density of a general scalar
field $p(\phi, X)$, where
$X=-g^{\mu\nu}\partial_\mu\phi\partial_\nu\phi/2$ is the kinetic
energy term. Note that $p(\phi, X)$ is a general function of $\phi$
and $X$, and we have used a sign notation $(-, +, +, +)$.
Identifying the energy momentum tensor of the scalar field with that
of a perfect fluid, we can easily derive the energy density,
$\rho_{\rm de}=2Xp_X-p$, where $p_X=\partial p/\partial X$. Thus, in
a spatially flat FRW universe, the dynamic equations for the scalar
field are
\begin{equation}
3H^2=\rho_{\rm m}+2Xp_X-p,\label{hsqr}
\end{equation}
\begin{equation}
2\dot{H}=-\rho_{\rm m}-2Xp_X,\label{hdot}
\end{equation}
where $X=\dot{\phi}^2/2$ in the cosmological context. Introducing a
dimensionless quantity
\begin{equation}
r\equiv H^2/H_0^2,
\end{equation}
we find from Eqs.(\ref{hsqr}) and (\ref{hdot}) that
\begin{equation}
p=[(1+z)r'-3r]H_0^2,\label{p}
\end{equation}
\begin{equation}
\phi'^2p_X={r'-3\Omega_{\rm m}^0(1+z)^2\over r(1+z)},\label{px}
\end{equation}
where prime denotes a derivative with respect to $z$. The equation
of state for dark energy is given by
\begin{equation}
w={p\over \dot{\phi}^2 p_X-p}={(1+z)r'-3r\over 3r-3\Omega_{\rm
m}^0(1+z)^3}.
\end{equation}
Next, if we establish a correspondence between the holographic
vacuum energy and the scalar field dark energy, we should choose a
scalar field model in which crossing the cosmological-constant
boundary is possible. So, let us consider the generalized ghost
condensate model proposed in Ref.\cite{general}, with the Lagrangian
density
\begin{equation}
p=-X+h(\phi)X^2,
\end{equation}
where $h(\phi)$ is a function in terms of $\phi$. Dilatonic ghost
condensate model \cite{ghost2} corresponds to a choice
$h(\phi)=ce^{\lambda\phi}$. From Eqs. (\ref{p}) and (\ref{px}) we
obtain
\begin{equation}
\phi'^2={12r-3(1+z)r'-3\Omega_{\rm m}^0(1+z)^3\over r(1+z)^2},
\end{equation}
\begin{equation}
h(\phi)={6(2(1+z)r'-6r+r(1+z)^2\phi'^2)\over
r^2(1+z)^4\phi'^4}\rho_{\rm c0}^{-1},
\end{equation}
where $\rho_{\rm c0}=3H_0^2$ represents the present critical density
of the universe. The generalized ghost condensate describes the
holographic vacuum energy, provided that
\begin{equation}
r={\Omega_{\rm m}^0(1+z)^3\over 1-\Omega_{\rm de}},
\end{equation}
\begin{equation}
r'={\Omega_{\rm m}^0(1+z)^2\over 1-\Omega_{\rm
de}}\left[3-\Omega_{\rm de}\left(1+{c\over 2}\sqrt{\Omega_{\rm
de}}\right)\right],
\end{equation}
where $\Omega_{\rm de}$ satisfies the deferential equation
(\ref{deq}).

%%%%%%%%%%%%%%%%%%%%%%%%%%%%%%%%%%%%%%%%%%%%%%%%%%%%%%%%%%%%%%%%%%
\begin{figure}[htbp]
\begin{center}
\includegraphics[scale=1.15]{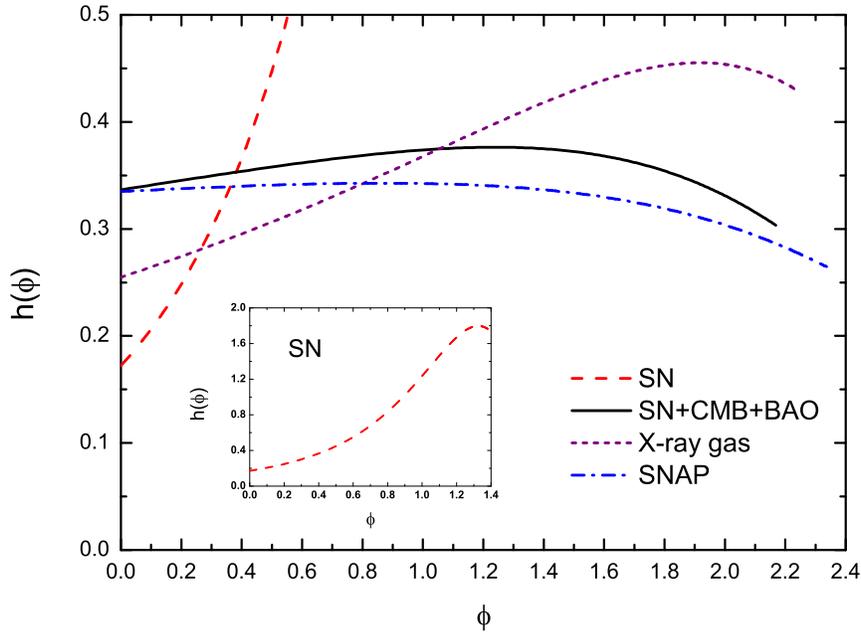}
\caption[]{\small Reconstruction of the generalized ghost condensate
model according to the holographic dark energy scenario. In this
plot, we show the cases of function $h(\phi)$, in unit of $\rho_{\rm
c0}^{-1}$, corresponding to the best fit results of SN, SN+CMB+BAO,
X-ray gas and SNAP, respectively.}
\end{center}
\end{figure}
%%%%%%%%%%%%%%%%%%%%%%%%%%%%%%%%%%%%%%%%%%%%%%%%%%%%%%%%%%%%%%%%%%%

%%%%%%%%%%%%%%%%%%%%%%%%%%%%%%%%%%%%%%%%%%%%%%%%%%%%%%%%%%%%%%%%%%
\begin{figure}[htbp]
\begin{center}
\includegraphics[scale=1.2]{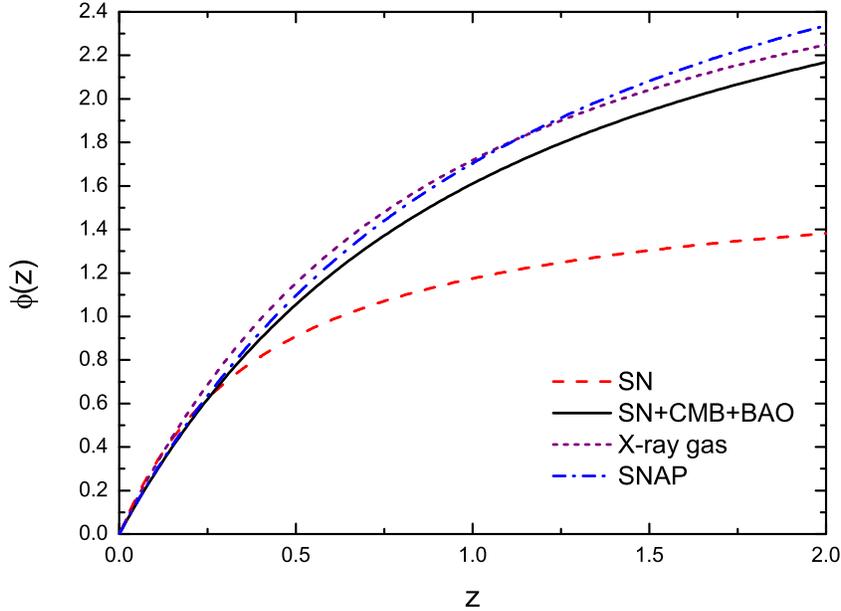}
\caption[]{\small Reconstruction of the generalized ghost condensate
model according to the holographic dark energy scenario. In this
plot, we show the evolutions of the scalar field $\phi(z)$,
corresponding to the best fit results of SN, SN+CMB+BAO, X-ray gas
and SNAP, respectively.}
\end{center}
\end{figure}
%%%%%%%%%%%%%%%%%%%%%%%%%%%%%%%%%%%%%%%%%%%%%%%%%%%%%%%%%%%%%%%%%%%

The reconstruction for $h(\phi)$ is plotted in Fig.2, using the
best-fit values of $c$ and $\Omega_{\rm m}^0$ from the observational
data analyses of SN Ia, SN+CMB+BAO, X-ray gas and SNAP (simulation),
respectively. The crossing of the cosmological-constant boundary
corresponds to $hX=1/2$. The system can enter the phantom region
($hX <1/2$) without discontinuous behavior of $h$ and $X$. In
addition, the evolution of the scalar field $\phi(z)$ is also
determined by the reconstruction program, see Fig.3. It should be
mentioned that the reconstruction of the generalized ghost
condensate model has been carried out in Ref.\cite{general} by using
the best-fit results of the parametrization for the Hubble parameter
$r(x)=\Omega_{\rm m}^0x^3+A_0+A_1x+A_2x^2$, where $x=1+z$ and
$A_0=1-A_1-A_2-\Omega_{\rm m}^0$, from the SN Gold dataset
\cite{cross1}. Our reconstruction result is consistent with that of
Ref.\cite{general}, except for the $c=0.21$ case (since the SN fit
result for the holographic dark energy model is not reasonable, for
details see Ref.\cite{obs1}). The future high-precision observations
are expected to determine the value of $c$ and the functional form
of $h(\phi)$ more accurately.

In conclusion, we suggest in this paper a correspondence between the
holographic dark energy scenario and a scalar field dark energy
model. We adopt the viewpoint of that the scalar field models of
dark energy are effective theories of an underlying theory of dark
energy. The underlying theory, though has not been achieved
presently, is presumed to possess some features of a quantum gravity
theory, which can be explored speculatively by taking into account
the holographic principle of quantum gravity theory. Consequently,
the vacuum energy acquires the dynamical property when imposing the
holographic principle. Moreover, the current available observational
data imply that the holographic vacuum energy behaves as
quintom-type dark energy, i.e. the equation-of-state of dark energy
crosses the cosmological-constant boundary $w=-1$ during the
evolution history. If we regard the scalar field model as an
effective description of such a theory (holographic vacuum), we
should be capable of using the scalar field model to mimic the
evolving behavior of the dynamical vacuum energy and reconstructing
this scalar field model according to the fits of the observational
dataset. We find the generalized ghost condensate model is a good
choice for depicting the holographic vacuum energy, since it can
easily realize the quintom behavior. We thus reconstructed the
function $h(\phi)$ of the generalized ghost condensate model using
the best-fit results of the observational data. We hope that the
future high precision observations (e.g. SNAP) may be capable of
determining the fine property of the dark energy and consequently
reveal some significant features of the underlying theory of dark
energy.

%\newpage

\section*{Acknowledgements}

The author would like to thank Hui Li, Miao Li, Yi Wang, Jingfei
Zhang and Yi Zhang for useful discussions. This work was supported
in part by the Natural Science Foundation of China.

%%%%%%%%%%%%%%%%%%%%%%%%%%%%%%%%%%%%%%%

%\section*{References}


\begin{thebibliography}{99}


%\cite{LSS}
\bibitem{LSS}
  M.~Tegmark {\it et al.}  [SDSS Collaboration],
  %``Cosmological parameters from SDSS and WMAP,''
  Phys.\ Rev.\ D {\bf 69}, 103501 (2004)
  [astro-ph/0310723];\\
  %%CITATION = ASTRO-PH 0310723;%%
  %\cite{Abazajian:2004aj}
%\bibitem{Abazajian:2004aj}
  K.~Abazajian {\it et al.}  [SDSS Collaboration],
  %``The Second Data Release of the Sloan Digital Sky Survey,''
  Astron.\ J.\  {\bf 128}, 502 (2004)
  [astro-ph/0403325];\\
  %%CITATION = ASTRO-PH 0403325;%%
%\cite{Abazajian:2004it}
%\bibitem{Abazajian:2004it}
  K.~Abazajian {\it et al.}  [SDSS Collaboration],
  %``The Third Data Release of the Sloan Digital Sky Survey,''
  Astron.\ J.\  {\bf 129}, 1755 (2005)
  [astro-ph/0410239].
  %%CITATION = ASTRO-PH 0410239;%%

%\cite{SN}
\bibitem{SN}
  A.~G.~Riess {\it et al.}  [Supernova Search Team Collaboration],
  %``Observational Evidence from Supernovae for an Accelerating Universe and a
  %Cosmological Constant,''
  Astron.\ J.\  {\bf 116}, 1009 (1998)
  [astro-ph/9805201];\\
  %%CITATION = ASTRO-PH 9805201;%%
%\cite{Perlmutter:1998np}
%\bibitem{Perlmutter:1998np}
  S.~Perlmutter {\it et al.}  [Supernova Cosmology Project Collaboration],
  %``Measurements of Omega and Lambda from 42 High-Redshift Supernovae,''
  Astrophys.\ J.\  {\bf 517}, 565 (1999)
  [astro-ph/9812133];\\
  %%CITATION = ASTRO-PH 9812133;%%
%\cite{Riess:2004nr}
%\cite{Astier:2005qq}
%\bibitem{Astier:2005qq}
  P.~Astier {\it et al.},
   %``The Supernova Legacy Survey: Measurement of Omega_M, Omega_Lambda and w
  %from the First Year Data Set,''
  Astron.\ Astrophys.\  {\bf 447}, 31 (2006)
  [astro-ph/0510447].
  %%CITATION = ASTRO-PH 0510447;%%



%\cite{CMB}
\bibitem{CMB}
  D.~N.~Spergel {\it et al.}  [WMAP Collaboration],
  %``First Year Wilkinson Microwave Anisotropy Probe (WMAP) Observations:
  %Determination of Cosmological Parameters,''
  Astrophys.\ J.\ Suppl.\  {\bf 148}, 175 (2003)
  [astro-ph/0302209];\\
  %%CITATION = ASTRO-PH 0302209;%%
%\cite{Spergel:2006hy}
%\bibitem{Spergel:2006hy}
  D.~N.~Spergel {\it et al.},
  %``Wilkinson Microwave Anisotropy Probe (WMAP) three year results:
  %Implications for cosmology,''
  astro-ph/0603449.
  %%CITATION = ASTRO-PH 0603449;%%

\bibitem{Einstein:1917} A. Einstein, Sitzungsber. K. Preuss.
Akad. Wiss. 142 (1917) [Einglish translation in {\it The Principle
of Relativity} (Dover, New York, 1952), p. 177].

\bibitem{cc}S.~Weinberg,
%``The Cosmological Constant Problem,''
Rev.\ Mod.\ Phys.\  {\bf 61} 1 (1989);\\
%\bibitem{Sahni:1999ijmpd}
  V.~Sahni and A.~A.~Starobinsky,
  %``The Case for a Positive Cosmological Lambda-term,''
  Int.\ J.\ Mod.\ Phys.\ D {\bf 9}, 373 (2000)
  [astro-ph/9904398];\\
  %%CITATION = ASTRO-PH 9904398;%%
%\bibitem{Carroll:2000}
S.~M.~Carroll,
%"The Cosmological Constant,"%
Living\ Rev.\ Rel.\ {\bf 4} 1 (2001) [astro-ph/0004075];\\
%\bibitem{Peebles:2003rmp}
P.~J.~E.~Peebles and B.~Ratra,
%``The cosmological constant and dark energy,''
Rev.\ Mod.\ Phys.\  {\bf 75} 559 (2003) [astro-ph/0207347];\\
%\bibitem{Padmanabhan:2003pr}
T.~Padmanabhan,
%``Cosmological constant: The weight of the vacuum,''
Phys.\ Rept.\  {\bf 380} 235 (2003) [hep-th/0212290].

\bibitem{coincidence} P.~J.~Steinhardt, in {\it Critical Problems in
Physics}, edited by V.~L.~Fitch and D.~R.~Marlow (Princeton
University Press, Princeton, NJ, 1997).

%\cite{quintessence}
\bibitem{quintessence}
  P.~J.~E.~Peebles and B.~Ratra,
  %``Cosmology With A Time Variable Cosmological 'Constant',''
  Astrophys.\ J.\  {\bf 325} L17 (1988);\\
  %%CITATION = ASJOA,325,L17;%%
%\cite{Ratra:1987rm}
%\bibitem{Ratra:1987rm}
  B.~Ratra and P.~J.~E.~Peebles,
  %``Cosmological Consequences Of A Rolling Homogeneous Scalar Field,''
  Phys.\ Rev.\ D {\bf 37} 3406 (1988);\\
  %%CITATION = PHRVA,D37,3406;%%
%\cite{Wetterich:1987fm}
%\bibitem{Wetterich:1987fm}
  C.~Wetterich,
  %``Cosmology And The Fate Of Dilatation Symmetry,''
  Nucl.\ Phys.\ B {\bf 302} 668 (1988);\\
  %%CITATION = NUPHA,B302,668;%%
%\cite{Frieman:1995pm}
%\bibitem{Frieman:1995pm}
  J.~A.~Frieman, C.~T.~Hill, A.~Stebbins and I.~Waga,
  %``Cosmology with ultralight pseudo Nambu-Goldstone bosons,''
  Phys.\ Rev.\ Lett.\  {\bf 75}, 2077 (1995)
  [astro-ph/9505060];\\
  %%CITATION = ASTRO-PH 9505060;%%
%\cite{Turner:1998ex}
%\bibitem{Turner:1998ex}
  M.~S.~Turner and M.~J.~White,
  %``CDM Models with a Smooth Component,''
  Phys.\ Rev.\ D {\bf 56}, 4439 (1997)
  [astro-ph/9701138];\\
  %%CITATION = ASTRO-PH 9701138;%%
%\cite{Caldwell:1997ii}
%\bibitem{Caldwell:1997ii}
  R.~R.~Caldwell, R.~Dave and P.~J.~Steinhardt,
  %``Cosmological Imprint of an Energy Component with General
  %Equation-of-State,''
  Phys.\ Rev.\ Lett.\  {\bf 80}, 1582 (1998)
  [astro-ph/9708069];\\
  %%CITATION = ASTRO-PH 9708069;%%
%\cite{Liddle:1998xm}
%\bibitem{Liddle:1998xm}
  A.~R.~Liddle and R.~J.~Scherrer,
  %``A classification of scalar field potentials with cosmological scaling
  %solutions,''
  Phys.\ Rev.\ D {\bf 59}, 023509 (1999)
  [astro-ph/9809272];\\
  %%CITATION = ASTRO-PH 9809272;%%
%\cite{tracker}
%\bibitem{tracker}
  I.~Zlatev, L.~M.~Wang and P.~J.~Steinhardt,
  %``Quintessence, Cosmic Coincidence, and the Cosmological Constant,''
  Phys.\ Rev.\ Lett.\  {\bf 82}, 896 (1999)
  [astro-ph/9807002];\\
  %%CITATION = ASTRO-PH 9807002;%%
%\cite{Steinhardt:1999nw}
%\bibitem{Steinhardt:1999nw}
  P.~J.~Steinhardt, L.~M.~Wang and I.~Zlatev,
  %``Cosmological tracking solutions,''
  Phys.\ Rev.\ D {\bf 59}, 123504 (1999)
  [astro-ph/9812313].
  %%CITATION = ASTRO-PH 9812313;%%

%\cite{kessence}
\bibitem{kessence}
  C.~Armendariz-Picon, V.~F.~Mukhanov and P.~J.~Steinhardt,
   %``A dynamical solution to the problem of a small cosmological constant  and
  %late-time cosmic acceleration,''
  Phys.\ Rev.\ Lett.\  {\bf 85}, 4438 (2000)
  [astro-ph/0004134];\\
  %%CITATION = ASTRO-PH 0004134;%%
%\cite{Armendariz-Picon:2000ah}
%\bibitem{Armendariz-Picon:2000ah}
  C.~Armendariz-Picon, V.~F.~Mukhanov and P.~J.~Steinhardt,
  %``Essentials of k-essence,''
  Phys.\ Rev.\ D {\bf 63}, 103510 (2001)
  [astro-ph/0006373].
  %%CITATION = ASTRO-PH 0006373;%%

%\cite{tachyon}
\bibitem{tachyon}
  A.~Sen,
  %``Tachyon matter,''
  JHEP {\bf 0207}, 065 (2002)
  [hep-th/0203265];\\
  %%CITATION = HEP-TH 0203265;%%
%\cite{Padmanabhan:2002cp}
%\bibitem{Padmanabhan:2002cp}
  T.~Padmanabhan,
  %``Accelerated expansion of the universe driven by tachyonic matter,''
  Phys.\ Rev.\ D {\bf 66}, 021301 (2002)
  [hep-th/0204150].
  %%CITATION = HEP-TH 0204150;%%

%\cite{phantom}
\bibitem{phantom}
  R.~R.~Caldwell,
  %``A Phantom Menace?,''
  Phys.\ Lett.\ B {\bf 545}, 23 (2002)
  [astro-ph/9908168];\\
  %%CITATION = ASTRO-PH 9908168;%%
%\cite{Caldwell:2003vq}
%\bibitem{Caldwell:2003vq}
  R.~R.~Caldwell, M.~Kamionkowski and N.~N.~Weinberg,
  %``Phantom Energy and Cosmic Doomsday,''
  Phys.\ Rev.\ Lett.\  {\bf 91}, 071301 (2003)
  [astro-ph/0302506].
  %%CITATION = ASTRO-PH 0302506;%%

%\cite{ghost}
\bibitem{ghost1}
  N.~Arkani-Hamed, H.~C.~Cheng, M.~A.~Luty and S.~Mukohyama,
  %``Ghost condensation and a consistent infrared modification of gravity,''
  JHEP {\bf 0405}, 074 (2004)
  [hep-th/0312099].
  %%CITATION = HEP-TH 0312099;%%


\bibitem{ghost2}
  F.~Piazza and S.~Tsujikawa,
  %``Dilatonic ghost condensate as dark energy,''
  JCAP {\bf 0407}, 004 (2004)
  [hep-th/0405054].
  %%CITATION = HEP-TH 0405054;%%
%\cite{Krause:2004bu}


%\cite{quintom}
\bibitem{quintom}
  B.~Feng, X.~L.~Wang and X.~M.~Zhang,
  %``Dark Energy Constraints from the Cosmic Age and Supernova,''
  Phys.\ Lett.\ B {\bf 607}, 35 (2005)
  [astro-ph/0404224];\\
  %%CITATION = ASTRO-PH 0404224;%%
%\cite{Feng:2004ff}
%\bibitem{Feng:2004ff}
  B.~Feng, M.~Li, Y.~S.~Piao and X.~M.~Zhang,
  %``Oscillating quintom and the recurrent universe,''
  Phys.\ Lett.\ B {\bf 634}, 101 (2006)
  [astro-ph/0407432];\\
  %%CITATION = ASTRO-PH 0407432;%%
%\cite{Guo:2004fq}
%\bibitem{Guo:2004fq}
  Z.~K.~Guo, Y.~S.~Piao, X.~M.~Zhang and Y.~Z.~Zhang,
  %``Cosmological evolution of a quintom model of dark energy,''
  Phys.\ Lett.\ B {\bf 608}, 177 (2005)
  [astro-ph/0410654];\\
  %%CITATION = ASTRO-PH 0410654;%%
%\cite{Zhang:2005kj}
%\bibitem{Zhang:2005kj}
  X.~Zhang,
  %``An interacting two-fluid scenario for quintom dark energy,''
  Commun.\ Theor.\ Phys.\  {\bf 44}, 762 (2005);\\
  %%CITATION = CTPMD,44,762;%%
%\cite{Zhang:2005eg}
%\bibitem{Zhang:2005eg}
  X.~F.~Zhang, H.~Li, Y.~S.~Piao and X.~M.~Zhang,
  %``Two-field models of dark energy with equation of state across -1,''
  Mod.\ Phys.\ Lett.\ A {\bf 21}, 231 (2006)
  [astro-ph/0501652];\\
  %%CITATION = ASTRO-PH 0501652;%%
%\cite{Wei:2005nw}
%\bibitem{Wei:2005nw}
  H.~Wei, R.~G.~Cai and D.~F.~Zeng,
  %``Hessence: A new view of quintom dark energy,''
  Class.\ Quant.\ Grav.\  {\bf 22}, 3189 (2005)
  [hep-th/0501160];\\
  %%CITATION = HEP-TH 0501160;%%
%\cite{Li:2005fm}
%\bibitem{Li:2005fm}
  M.~Z.~Li, B.~Feng and X.~M.~Zhang,
  %``A single scalar field model of dark energy with equation of state  crossing
  %-1,''
  JCAP {\bf 0512}, 002 (2005)
  [hep-ph/0503268];\\
  %%CITATION = HEP-PH 0503268;%%
  % according to Vikman's comment
  A. Anisimov, E. Babichev and A. Vikman,
  % ``B-inflation,''
  JCAP {\bf 0506}, 006 (2005)
  [astro-ph/0504560];\\
%\cite{Wei:2005si}
%\bibitem{Wei:2005si}
  H.~Wei and R.~G.~Cai,
  %``A note on crossing the phantom divide in hybrid dark energy model,''
  Phys.\ Lett.\ B {\bf 634}, 9 (2006)
  [astro-ph/0512018];\\
  %%CITATION = ASTRO-PH 0512018;%%
%\cite{Wei:2006tn}
%\bibitem{Wei:2006tn}
  H.~Wei and R.~G.~Cai,
  %``Interacting vector-like dark energy, the first and second cosmological
  %coincidence problems,''
  Phys.\ Rev.\ D {\bf 73}, 083002 (2006)
  [astro-ph/0603052];\\
  %%CITATION = ASTRO-PH 0603052;%%
  %\cite{Guo:2006pc}
%\bibitem{Guo:2006pc}
  Z.~K.~Guo, Y.~S.~Piao, X.~Zhang and Y.~Z.~Zhang,
  %``Two-field quintom models in the w-w' plane,''
  astro-ph/0608165;\\
  %%CITATION = ASTRO-PH 0608165;%%
%\cite{Cai:2006dm}
%\bibitem{Cai:2006dm}
  Y.~F.~Cai, H.~Li, Y.~S.~Piao and X.~M.~Zhang,
  %``Cosmic duality in quintom universe,''
  gr-qc/0609039.
  %%CITATION = GR-QC 0609039;%%




%\cite{Witten:2000zk}
\bibitem{Witten:2000zk}
  E.~Witten,
  %``The cosmological constant from the viewpoint of string theory,''
  hep-ph/0002297.
  %%CITATION = HEP-PH 0002297;%%

%\cite{Cohen:1998zx}
\bibitem{Cohen:1998zx}
  A.~G.~Cohen, D.~B.~Kaplan and A.~E.~Nelson,
  %``Effective field theory, black holes, and the cosmological constant,''
  Phys.\ Rev.\ Lett.\  {\bf 82}, 4971 (1999)
  [hep-th/9803132].
  %%CITATION = HEP-TH 9803132;%%

%\cite{Horava:2000tb}
\bibitem{Horava:2000tb}
  P.~Horava and D.~Minic,
  %``Probable values of the cosmological constant in a holographic theory,''
  Phys.\ Rev.\ Lett.\  {\bf 85}, 1610 (2000)
  [hep-th/0001145];\\
  %%CITATION = HEP-TH 0001145;%%
%\cite{Thomas:2002pq}
%\bibitem{Thomas:2002pq}
  S.~D.~Thomas,
  %``Holography stabilizes the vacuum energy,''
  Phys.\ Rev.\ Lett.\  {\bf 89}, 081301 (2002).
  %%CITATION = PRLTA,89,081301;%%

%\cite{Hsu:2004ri}
\bibitem{Hsu:2004ri}
  S.~D.~H.~Hsu,
  %``Entropy bounds and dark energy,''
  Phys.\ Lett.\ B {\bf 594}, 13 (2004)
  [hep-th/0403052].
  %%CITATION = HEP-TH 0403052;%%

%\cite{Li:2004rb}
\bibitem{Li:2004rb}
  M.~Li,
  %``A model of holographic dark energy,''
  Phys.\ Lett.\ B {\bf 603}, 1 (2004)
  [hep-th/0403127].
  %%CITATION = HEP-TH 0403127;%%

%\cite{holoprin}
\bibitem{holoprin}
  G.~'t Hooft,
  %``Dimensional reduction in quantum gravity,''
  gr-qc/9310026;\\
  %%CITATION = GR-QC 9310026;%%
%\cite{Susskind:1994vu}
%\bibitem{Susskind:1994vu}
  L.~Susskind,
  %``The World as a hologram,''
  J.\ Math.\ Phys.\  {\bf 36}, 6377 (1995)
  [hep-th/9409089].
  %%CITATION = HEP-TH 9409089;%%


\bibitem{bh}
J. D. Bekenstein, Phys. Rev. D {\bf7} (1973) 2333;\\
J. D. Bekenstein, Phys. Rev. D {\bf9} (1974) 3292; J. D. Bekenstein,
Phys. Rev. D {\bf23} (1981) 287;\\
J. D. Bekenstein, Phys. Rev. D {\bf49} (1994) 1912;\\
S. W. Hawking, Commun. Math. Phys. {\bf43} (1975) 199;\\
S. W. Hawking, Phys. Rev. D {\bf13} (1976) 191.







\bibitem{obs1}
  X.~Zhang and F.~Q.~Wu,
  %``Constraints on holographic dark energy from Type Ia supernova
  %observations,''
  Phys.\ Rev.\ D {\bf 72}, 043524 (2005)
  [astro-ph/0506310].
  %%CITATION = ASTRO-PH 0506310;%%


\bibitem{obs2}
  Z.~Chang, F.~Q.~Wu and X.~Zhang,
  %``Constraints on holographic dark energy from X-ray gas mass fraction of
  %galaxy clusters,''
  Phys.\ Lett.\ B {\bf 633}, 14 (2006)
  [astro-ph/0509531].
  %%CITATION = ASTRO-PH 0509531;%%

\bibitem{obs3}
  Q.~G.~Huang and Y.~G.~Gong,
  %``Supernova constraints on a holographic dark energy model,''
  JCAP {\bf 0408}, 006 (2004)
  [astro-ph/0403590];\\
  %%CITATION = ASTRO-PH 0403590;%%
%\bibitem{obs4}
  K.~Enqvist, S.~Hannestad and M.~S.~Sloth,
  %``Searching for a holographic connection between dark energy and the  low-l
  %CMB multipoles,''
  JCAP {\bf 0502} 004 (2005)
  [astro-ph/0409275];\\
  %%CITATION = ASTRO-PH 0409275;%%
%\cite{Shen:2004ck}
%\bibitem{Shen:2004ck}
  J.~Shen, B.~Wang, E.~Abdalla and R.~K.~Su,
  %``Constraints on the dark energy from the holographic connection to the
  %small l CMB suppression,''
  Phys.\ Lett.\ B {\bf 609} 200 (2005)
  [hep-th/0412227];\\
  %%CITATION = HEP-TH 0412227;%%
%\cite{Kao:2005xp}
%\bibitem{Kao:2005xp}
  H.~C.~Kao, W.~L.~Lee and F.~L.~Lin,
  %``CMB constraints on the holographic dark energy model,''
  Phys.\ Rev.\ D {\bf 71} 123518 (2005)
  [astro-ph/0501487].
  %%CITATION = ASTRO-PH 0501487;%%


%\cite{holoext}
\bibitem{holoext}
  Q.~G.~Huang and M.~Li,
  %``The holographic dark energy in a non-flat universe,''
  JCAP {\bf 0408}, 013 (2004)
  [astro-ph/0404229];\\
  %%CITATION = ASTRO-PH 0404229;%%
%\cite{Ito:2004qi}
%\bibitem{Ito:2004qi}
%  M.~Ito,
  %``Holographic dark energy model with non-minimal coupling,''
%  Europhys.\ Lett.\  {\bf 71}, 712 (2005)
%  [hep-th/0405281];\\
  %%CITATION = HEP-TH 0405281;%%
  %\cite{Enqvist:2004xv}
%\bibitem{Enqvist:2004xv}
  K.~Enqvist and M.~S.~Sloth,
  %``A CMB / dark energy cosmic duality,''
  Phys.\ Rev.\ Lett.\  {\bf 93}, 221302 (2004)
  [hep-th/0406019];\\
  %%CITATION = HEP-TH 0406019;%%
%\cite{Ke:2004nw}
%\bibitem{Ke:2004nw}
  K.~Ke and M.~Li,
  %``Cardy-Verlinde formula and holographic dark energy,''
  Phys.\ Lett.\ B {\bf 606}, 173 (2005)
  [hep-th/0407056];\\
  %%CITATION = HEP-TH 0407056;%%
%\cite{Huang:2004mx}
%\bibitem{Huang:2004mx}
  Q.~G.~Huang and M.~Li,
  %``Anthropic principle favors the holographic dark energy,''
  JCAP {\bf 0503}, 001 (2005)
  [hep-th/0410095];\\
  %%CITATION = HEP-TH 0410095;%%
%\cite{Zhang:2005yz}
%\bibitem{Zhang:2005yz}
  X.~Zhang,
  %``Statefinder diagnostic for holographic dark energy model,''
  Int.\ J.\ Mod.\ Phys.\ D {\bf 14}, 1597 (2005)
  [astro-ph/0504586];\\
  %%CITATION = ASTRO-PH 0504586;%%
%\cite{Pavon:2005yx}
%\bibitem{Pavon:2005yx}
  D.~Pavon and W.~Zimdahl,
  %``Holographic dark energy and cosmic coincidence,''
  Phys.\ Lett.\ B {\bf 628}, 206 (2005)
  [gr-qc/0505020];\\
  %%CITATION = GR-QC 0505020;%%
%\cite{Wang:2005jx}
%\bibitem{Wang:2005jx}
  B.~Wang, Y.~Gong and E.~Abdalla,
  %``Transition of the dark energy equation of state in an interacting
  %holographic dark energy model,''
  Phys.\ Lett.\ B {\bf 624}, 141 (2005)
  [hep-th/0506069];\\
  %%CITATION = HEP-TH 0506069;%%
%\cite{Kim:2005at}
%\bibitem{Kim:2005at}
  H.~Kim, H.~W.~Lee and Y.~S.~Myung,
  %``Equation of state for an interacting holographic dark energy model,''
  Phys.\ Lett.\ B {\bf 632}, 605 (2006)
  [gr-qc/0509040];\\
  %%CITATION = GR-QC 0509040;%%
  %\cite{Nojiri:2005pu}
%\bibitem{Nojiri:2005pu}
  S.~Nojiri and S.~D.~Odintsov,
  % ``Unifying phantom inflation with late-time acceleration: Scalar
  % phantom-non-phantom transition model and generalized holographic dark
  %energy,''
  Gen.\ Rel.\ Grav.\  {\bf 38}, 1285 (2006)
  [hep-th/0506212];\\
  %%CITATION = HEP-TH 0506212;%%
%\cite{Elizalde:2005ju}
%\bibitem{Elizalde:2005ju}
  E.~Elizalde, S.~Nojiri, S.~D.~Odintsov and P.~Wang,
   %``Dark energy: Vacuum fluctuations, the effective phantom phase, and
  %holography,''
  Phys.\ Rev.\ D {\bf 71}, 103504 (2005)
  [hep-th/0502082];\\
  %%CITATION = HEP-TH 0502082;%%
%\cite{Hu:2006ar}
%\bibitem{Hu:2006ar}
  B.~Hu and Y.~Ling,
  %``Interacting dark energy, holographic principle and coincidence problem,''
  Phys.\ Rev.\ D {\bf 73}, 123510 (2006)
  [hep-th/0601093];\\
  %%CITATION = HEP-TH 0601093;%%
%\cite{Li:2006ci}
%\bibitem{Li:2006ci}
  H.~Li, Z.~K.~Guo and Y.~Z.~Zhang,
  %``A tracker solution for a holographic dark energy model,''
  Int.\ J.\ Mod.\ Phys.\ D {\bf 15}, 869 (2006)
  [astro-ph/0602521];\\
  %%CITATION = ASTRO-PH 0602521;%%
  %\cite{Setare:2006vz}
%\bibitem{Setare:2006vz}
  M.~R.~Setare and S.~Shafei,
   %``The holographic model of dark energy and thermodynamics of non-flat
  %accelerated expanding universe,''
  JCAP {\bf 0609}, 011 (2006)
  [gr-qc/0606103];\\
  %%CITATION = GR-QC 0606103;%%
%\cite{Setare:2006wh}
%\bibitem{Setare:2006wh}
  M.~R.~Setare,
  %``Interacting holographic dark energy model in non-flat universe,''
  Phys.\ Lett.\ B {\bf 642}, 1 (2006)
  [hep-th/0609069];\\
  %%CITATION = HEP-TH 0609069;%%
%\cite{Setare:2006pj}
%\bibitem{Setare:2006pj}
  M.~R.~Setare,
  %``Bulk-brane interaction and holographic dark energy,''
  hep-th/0609104;\\
  %%CITATION = HEP-TH 0609104;%%
%\cite{Sadjadi:2006qb}
%\bibitem{Sadjadi:2006qb}
  H.~M.~Sadjadi and M.~Honardoost,
   %``Thermodynamics second law and omega = -1 crossing(s) in interacting
  %holographic dark energy model,''
  gr-qc/0609076.
  %%CITATION = GR-QC 0609076;%%











\bibitem{cross1}
  U.~Alam, V.~Sahni and A.~A.~Starobinsky,
  %``The case for dynamical dark energy revisited,''
  JCAP {\bf 0406}, 008 (2004)
  [astro-ph/0403687].
  %%CITATION = ASTRO-PH 0403687;%%


\bibitem{cross2}
  D.~Huterer and A.~Cooray,
  %``Uncorrelated Estimates of Dark Energy Evolution,''
  Phys.\ Rev.\ D {\bf 71}, 023506 (2005)
  [astro-ph/0404062].
  %%CITATION = ASTRO-PH 0404062;%%
%\cite{Zhao:2006bt}


\bibitem{Riess:2004nr}
  A.~G.~Riess {\it et al.}  [Supernova Search Team Collaboration],
  %``Type Ia Supernova Discoveries at z>1 From the Hubble Space Telescope:
  %Evidence for Past Deceleration and Constraints on Dark Energy Evolution,''
  Astrophys.\ J.\  {\bf 607}, 665 (2004)
  [astro-ph/0402512].
  %%CITATION = ASTRO-PH 0402512;%%



%\cite{BAO}
\bibitem{BAO}
  D.~J.~Eisenstein {\it et al.}  [SDSS Collaboration],
   %``Detection of the Baryon Acoustic Peak in the Large-Scale Correlation
  %Function of SDSS Luminous Red Galaxies,''
  Astrophys.\ J.\  {\bf 633}, 560 (2005)
  [astro-ph/0501171].
  %%CITATION = ASTRO-PH 0501171;%%



%\cite{nl}
\bibitem{nl}
   A. Vikman,
   % "Can dark energy evolve to the phantom?"
   Phys. Rev. D {\bf 71}, 023515 (2005)
   [astro-ph/0407107].

\bibitem{general}
S. Tsujikawa, Phys. Rev. D {\bf72}, 083512 (2005)
[astro-ph/0508542];\\
E. J. Copeland, M. Sami and S. Tsujikawa, hep-th/0603057.


%\cite{Saini:1999ba}
\bibitem{Saini:1999ba}
  T.~D.~Saini, S.~Raychaudhury, V.~Sahni and A.~A.~Starobinsky,
  %``Reconstructing the Cosmic Equation of State from Supernova distances,''
  Phys.\ Rev.\ Lett.\  {\bf 85}, 1162 (2000)
  [astro-ph/9910231].
  %%CITATION = ASTRO-PH 9910231;%%

\bibitem{simplescalar}
A. A. Starobinsky, JETP Lett. {\bf68}, 757 (1998) [Pis'ma Zh. Eksp.
Teor. Fiz. {\bf 68}, 721 (1998)] [astro-ph/9810431];\\
D. Huterer and M. S. Turner, Phys. Rev. D {\bf60}, 081301 (1999)
[astro-ph/9808133];\\
T. Nakamura and T. Chiba, Mon. Not. R. Astron. Soc. {\bf306}, 696
(1999) [astro-ph/9810447].





%\cite{Boisseau:2000pr}
\bibitem{Boisseau:2000pr}
  B.~Boisseau, G.~Esposito-Farese, D.~Polarski and A.~A.~Starobinsky,
  % ``Reconstruction of a scalar-tensor theory of gravity in an accelerating
  %universe,''
  Phys.\ Rev.\ Lett.\  {\bf 85}, 2236 (2000)
  [gr-qc/0001066].
  %%CITATION = GR-QC 0001066;%%

%\cite{Szydlowski:2003cf}
\bibitem{Szydlowski:2003cf}
  M.~Szydlowski and W.~Czaja,
  %``Particle-like description in quintessential cosmology,''
  Phys.\ Rev.\ D {\bf 69}, 083518 (2004)
  [gr-qc/0305033];\\
  %%CITATION = GR-QC 0305033;%%
%\cite{Szydlowski:2003fg}
%\bibitem{Szydlowski:2003fg}
  M.~Szydlowski and W.~Czaja,
  % ``Toward reconstruction of the dynamics of the Universe from distant type Ia
  %supernovae,''
  Phys.\ Rev.\ D {\bf 69}, 083507 (2004)
  [astro-ph/0309191].
  %%CITATION = ASTRO-PH 0309191;%%


%\cite{Zhang:2006av}
\bibitem{Zhang:2006av}
  X.~Zhang,
  %``Reconstructing holographic quintessence,''
  astro-ph/0604484.
  %%CITATION = ASTRO-PH 0604484;%%

%\cite{Guo:2005at}
\bibitem{Guo:2005at}
  Z.~K.~Guo, N.~Ohta and Y.~Z.~Zhang,
  %``Parametrization of quintessence and its potential,''
  Phys.\ Rev.\ D {\bf 72}, 023504 (2005)
  [astro-ph/0505253];\\
  %%CITATION = ASTRO-PH 0505253;%%
%\cite{Li:2006bx}
%\bibitem{Li:2006bx}
  H.~Li, Z.~K.~Guo and Y.~Z.~Zhang,
  %``Parametrization of K-essence and its kinetic term,''
  Mod.\ Phys.\ Lett.\ A {\bf 21}, 1683 (2006)
  [astro-ph/0601007];\\
  %\cite{Guo:2006ab}
%\bibitem{Guo:2006ab}
  Z.~K.~Guo, N.~Ohta and Y.~Z.~Zhang,
  %``Parametrizations of the dark energy density and scalar potentials,''
  astro-ph/0603109.
  %%CITATION = ASTRO-PH 0603109;%%

%\cite{Zhao:2006mp}
\bibitem{Zhao:2006mp}
  W.~Zhao and Y.~Zhang,
  %``The quintom models with state equation crossing -1,''
  Phys.\ Rev.\ D {\bf 73}, 123509 (2006)
  [astro-ph/0604460].
  %%CITATION = ASTRO-PH 0604460;%%
































\end{thebibliography}
\end{document}